\documentstyle[11pt]{article}
\newcounter{thanksnum}
\def\thanksnumber#1
{\setcounter{thanksnum}{\value{footnote}}\setcounter{footnote}{#1}%
                     \addtocounter{footnote}{-1}\footnotemark
                     \setcounter{footnote}{\value{thanksnum}}}
\def\newtheoremz#1{\@ifnextchar[{\@othmz{#1}}{\@nthmz{#1}}}

\def\@nthmz#1#2{%
\@ifnextchar[{\@xnthmz{#1}{#2}}{\@ynthmz{#1}{#2}}}

\def\@xnthmz#1#2[#3]{\expandafter\@ifdefinable\csname #1\endcsname
{\@definecounter{#1}\@addtoreset{#1}{#3}%
\expandafter\xdef\csname the#1\endcsname{\expandafter\noexpand
  \csname the#3\endcsname \@thmcountersepz \@thmcounterz{#1}}%
\global\@namedef{#1}{\@thmz{#1}{#2}}\global\@namedef{end#1}{\@endtheoremz}}}

\def\@ynthmz#1#2{\expandafter\@ifdefinable\csname #1\endcsname
{\@definecounter{#1}%
\expandafter\xdef\csname the#1\endcsname{\@thmcounterz{#1}}%
\global\@namedef{#1}{\@thm{#1}{#2}}\global\@namedef{end#1}{\@endtheoremz}}}

\def\@othmz#1[#2]#3{\expandafter\@ifdefinable\csname #1\endcsname
  {\global\@namedef{the#1}{\@nameuse{the#2}}%
\global\@namedef{#1}{\@thmz{#2}{#3}}%
\global\@namedef{end#1}{\@endtheoremz}}}

\def\@thmz#1#2{\refstepcounter
    {#1}\@ifnextchar[{\@ythmz{#1}{#2}}{\@xthmz{#1}{#2}}}

\def\@xthmz#1#2{\@begintheoremz{#2}{\csname the#1\endcsname}\ignorespaces}
\def\@ythmz#1#2[#3]{\@opargbegintheoremz{#2}{\csname
       the#1\endcsname}{#3}\ignorespaces}

\def\@thmcounterz#1{\noexpand\arabic{#1}}
\def\@thmcountersepz{.}
\def\@begintheoremz#1#2{ \trivlist \item[\hskip \labelsep{\bf #1\ #2}]}
\def\@opargbegintheoremz#1#2#3{ \trivlist
      \item[\hskip \labelsep{\bf #1\ #2\ (#3)}]}
\def\@endtheoremz{\endtrivlist}

\newtheorem{theorem}{Theorem}[section]
\newtheorem{lemma}{Lemma}[section]
\newtheorem{assumption}{Assumption}[section]
\newtheorem{proposition}{Proposition}[section]
\newtheorem{corollary}{Corollary}[section]
\newtheorem{condition}{Condition}[section]
\newtheorem{definition}{Definition}[section]

\def\defi{\stackrel{{\scriptscriptstyle \Delta}}{=}}

\def\a{\alpha}
\def\d{\delta}
\def\o{\omega}
\def\O{\Omega}

\def\Y{{\cal Y}}
\def\F{{\cal F}}
\def\w{\widehat}

\def\const{{\rm const\,}}

\def\R{{\bf R}}
\def\E{{\bf E}}
\def\P{{\bf P}}

\def\Z{{\cal Z}}

\def\b{\beta}
\def\s{\delta}
\def\g{\gamma}

\def\ww{\widetilde}

\def\t{\theta}
\def\oo{\bar}
\def\s{\sigma}

\def\p{\partial}
\def\G{\Gamma}

\newcommand{\be}{\begin{equation}}
\newcommand{\ee}{\end{equation}}
\newcommand{\bd}{\begin{displaymath}}
\newcommand{\ed}{\end{displaymath}}
\newcommand{\ba}{\begin{array}{ll}}
\newcommand{\ea}{\end{array}}

\def\T{{\cal T}}

\def\oo{\bar}
\def\Fo{\F^{R,r}}
\def\a{\alpha}
\title{
Optimal portfolio with unobservable market  parameters
and certainty equivalence principle}
\author{
Nikolai Dokuchaev}
 
\begin{document} \maketitle
\begin{abstract} We consider a multi-stock continuous time incomplete 
market model with random coefficients. We study
the  investment problem in the class of strategies which do not
use direct observations of the appreciation rates of the stocks,
but rather use historical stock  prices and an  a priory given
distribution of the appreciation rates.  An explicit solution
is found for case of power utilities and for a case when the
problem can be embedded to a Markovian setting. Some new
estimates and filters for the appreciation rates are given.
\par   {\bf Key words}: Optimal portfolio, continuous time market
model,  non-observable parameters, filters \par {\bf JEL classification}:
D52, D81, D84, G11  \par   {\bf Mathematical Subject Classification:} 49K45, 60G15, 93E20
\end{abstract}
\section{Introduction} The paper investigates an optimal
investment problem  for a  market consisting of a locally risk
free asset  and a finite number, $n$, of risky stocks.   It is
assumed that the vector of stock prices $S(t)$ evolves according
to an   It\^o stochastic differential   equation with a vector of
appreciation rates $a(t)$ and a volatility matrix $\s(t)$:
$$dS_i(t)=S_i(t)[a_i(t)\ dt + \sum_j \s_{ij}(t)\ dw_j(t)],\quad
i=1,...,n.$$ The problem goes back to Merton (1969), who found
strategies which solve the optimization problem in which      $\E
U(X(T))$ is to be maximized, where $X(T)$ represents   the wealth
at the final time $T$ and where $U(\cdot)$ is a utility function.
If the market parameters are observed, then the  optimal
strategies (i.e. current vector of stock holdings) are functions
of the current vector $(a(t),\s(t),S(t),X(t))$; see, e.g., survey
in Hakansson (1997) and Karatzas and Shreve (1998). However, in
practice, the process  $(a(t),\s(t))$ is not given directly and has to be estimated from observations of prices given some prior hypothesis about the market dynamics.  
Some attempts
have been made to construct winning strategies that are not using these parameters;
see, e.g., Dokuchaev and Savkin  (2002), Dokuchaev (2002, 2007).
However, the mainstream approach is to consider models where
  $a(t)$ and $\s(t)$ have to be estimated from historical
stock prices or some other observation process.     There are many
papers devoted to estimation of  $(a(t),\s(t))$, mainly based on
modifications of Kalman-Bucy filtering or the maximum likelihood
principle; see e.g. Lo (1988), Chen and Scott (1993), Pearson and
Sun (1994).
   Unfortunately, the process $a(t)$ is usually hard
to estimate in real-time markets,  because the drift term, $a(t)$,
is usually overshadowed by the  diffusion term, $\s(t)$. On the
other hand, $\s(t)$ can, in principle, be found from stock prices;
see (\ref{vol}) below. Thus, there remains the problem of
optimal investment with unobservable $a(t)$.  A popular tool for
this problem is the so-called  { ``certainty equivalence
principle'':}  {\it agents who know the solution of the optimal
investment  problem for the case of directly observable $a(t)$ can
solve the problem with unobservable $a(t)$ by substituting
$\E\{a(t)|S(\tau),\tau<t\}$} (see e.g. Gennotte (1986), Feldman (2007)).
Unfortunately, this principle does not hold in the general case of
non-log utilities (see  Kuwana (1995)). Note that this principle
is unrelated to the notion of ``certainty
equivalent value" to be found in the work of Frittelli (2000).
\par In fact, the problem is one of linear filtering. If $R_i(t)$
is the return on the $i$th stock, then $$dR(t)= a(t) dt +\s(t)
dw(t),$$ so the estimation of $a(t)$ given $\{R(\tau), \tau<t\}$
(or $\{S(\tau), \tau<t\}$), is a linear filtering problem. {\em If
$a(\cdot)$ is conditionally Gaussian,}  then the Kalman filter
provides the estimate which minimizes the  error in the mean
square sense. Indeed, in this case the conditional mean and
conditional variance of $a(t)$ given the past prices completely
describe the conditional distribution of $a(t)$, ${\cal
P}_{a(t)}(\,\cdot\,|S(\tau),\tau<t)$. In this setting  Williams
(1977), Detemple (1986), Dothan and Feldman (1986), Gennotte
(1986), Brennan (1998) solved the investment problem using the
Kalman-Bucy filter. This solution is optimal  in the class of
admissible strategies which are functions of the current
$\left(X(t),S(t),{\cal P}_{a(t)}(\cdot|S(\tau),\tau<t)\right)$.
But in general for non-Gaussian $a$, the optimal strategy based on
all historical prices does not lie in this set, hence their
approach does not give the optimal strategy. \par {Karatzas}
(1997), Karatzas and Zhao (1998), {Dokuchaev and Zhou} (2000),
Dokuchaev and Teo (2000) have obtained optimal portfolio
strategies in the class of strategies of the form
$\pi(t)=f(\{S(\tau):\tau<t\})$, where $f(\cdot)$ is a
deterministic function, when $a(t)$ is  random and unobservable,
but under the crucial condition that  $a$ and $\s$ are time
independent. This assumption ensures that the optimal wealth has
the form $X(t)=H(S(t),t)$, where $H(\cdot,\cdot)$ satisfies a
deterministic parabolic backward equation of dimension $n$, for
the market with $n$ stocks. Even if one accepts this restrictive
condition, the solution of the problem is difficult to realise in
practice for large $n$ (say,  $n>4$), since it is usually
difficult to solve the parabolic equation.  {Karatzas} (1997)
gives the explicit solution of a  goal achieving problem for the
case of one stock which has conditionally normal growth rate.
Karatzas and Zhao (2001)  use linear filtering (via martingales)
and dynamic programming to solve the problem  for a general
utility function with  $n>1$, $\s$ diagonal and constant, and with
random $a$ with known (non-Gaussian) distribution.  In {Dokuchaev
and Zhou} (2000), additional constraints on the terminal wealth
are added so that goal achieving problems are subsumed. Dokuchaev
and Teo (2000)  further generalize the constraints and utility
functions allowed. \par The restriction of the constant coefficients
is relaxed in three seminal papers: Karatzas and Xue (1991) and
Lakner (1995), (1998). Karatzas and Xue assumed that there are more
Brownian motions than stocks. They assume that $r$ and $\s$ are
adapted to the observable $S$. After projecting onto an
$n$-dimensional Brownian motion which generates the same
filtration as $S$, they obtain a reduced, completely observable
model; existence of an optimal portfolio follows, but the optimal
strategy is, as usual, defined only implicitly. Lakner (1995),
(1998) assumes that $S$ and $w$ have equal dimension (as we do),
and that $r$ and $\s$ are deterministic. This again guarantees
that the filtration of $S$ is Brownian. Results from filtering
theory give a representation of the optimal portfolio, which is
explicit in terms of a conditional expectation of a Malliavin
derivative when the $a_i$ are Ornstein-Uhlenbeck processes
independent of $w$. Zohar (2001) suggested an alternative approach
based on a Cameron-Martin formula for a special single stock model
when parameters are described by  Ornstein-Uhlenbeck process.
\par We also consider the optimal investment
problem with random, unobservable $a(t)$, and we allow the random
coefficients, $r,\s$, to depend on time. Our approach, as usual,
is to exhibit a claim which gives the optimal terminal wealth; the
replication strategy for this claim will then be the optimal
strategy. The replicating strategies for the very general models
were obtained as a a conditional expectation of a Malliavin
derivative. In the present paper we are trying alternative
approaches that can give a more explicit solution for some special
cases.\par
 We are targeting two
special cases: (i) when $U(x)=x^\d$; (ii) when $\w a(t)$ can be
presented as a  part of a diffusion Markov process, may be of a
higher dimension.\par
\par First for the log utility and some  power utilities, we can
compute the hedging portfolio directly with few restrictions on
$r,a,\s$ (at least in the log case). For these utilities it is
shown that the "certainty equivalence principle" can be
reformulated with the following correction: the "equivalence
filter" of $a(t)$ must be derived (in place of
$\E\{a(t)|S(\tau),\tau<t\}$). In general, it is neither
$\E\{a(t)|S(\tau),\tau<t\}$ nor any other function of ${\cal
P}_{a(t)}(\,\cdot\,|S(\tau),\tau<t)$. We show that for a general
prior distribution of $a(\cdot)$ and logarithmic utility, the
equivalence filter of $a(t)$ is in fact $\w a(t)\defi \E\{
a(t)|S(\tau),\tau<t\}$. If $a(t)$ is Gaussian, then of course this
is the Kalman-Bucy filter; this case was considered in Lakner (1995),
(1998), Dokuchaev (2005). 
\par Further, for the case of power
utility the equivalence filter is not a function of ${\cal
P}_{a(t)}(\,\cdot\,|S(\tau),\tau<t)$ even under the  Gaussian
assumption. However, we show that this estimate can be written as
a conditional expectation of $a(t)$ {\em under a new measure}.
Thus, under a Gaussian assumption on $a(t)$, the equivalence
filter can be obtained by a Kalman-Bucy filter, {\em but} with
some correction to the parameters. In other words, our result
gives new filters which (for Gaussian priors) are the classic
Kalman-Bucy filters  but with modified parameters. Cvitani\'c {\it
et all} (2002) presented the explicit optimal strategy for
non-observable parameters for Gaussian priors for
$U(x)=\d^{-1}x^\d$, but only for the case when $\d<0$. Our
approach also leads to  the explicit solution but for {\it
positive}  $\d$; in fact, we cover   only the case when
$\d=(l-1)/l$, $l=2,3,...$. \par
 The second class of considered problems includes problems which
can be embedded to a Markovian setting. Dokuchaev and Zhou (2001)
suggested to use linear parabolic equations to replicate the
optimal claim when the coefficients are constant in time; in that
case, the optimal claim  can be presented as a function of the
vector of stock prices at terminal time  (see also Dokuchaev and
Teo (2000)). We extended this approach to a more general model
that covers the cases when $\Theta$ is an Ornstein-Ulenbek
process, or when $\Theta$ is a finitely-valued Markov process;  in
the first case, the solution requires solving a {\it linear}
parabolic equation of dimension $n+2$ and in the second case,
solving a parabolic equation of dimension equal to the number of
possible values of the Markov process $\Theta$. Thus, we propose a
simpler method than
 dynamic programming: the {\it nonlinear} parabolic Bellman
equation is replaced for  a  linear parabolic equation. Note that
Sass and Haussmann (2003) solved a more general problem for the
case of parameters being driven by a finitely valued Markov chain,
but their solution  presents the replicating strategy as a
conditional expectation of a Malliavin derivative.
\par In Section 2 we collect notation and definitions, and we set
up the model. The problem is stated in Section 3, and in Section 4
a formula for the optimal claim  is presented in a very general
setting. In Section \ref{sec4'} the solution is detailed for
 some power utilities.  In Section \ref{sec6} we consider the
cases when the problem can  be embedded  to a   Markovian
setting.  The Appendix contains most of the proofs. \par
\section{The market model} On a given
probability space $(\O,\F,\P)$  satisfying the usual conditions,
consider a market model consisting of a locally risk free asset or
bank account with price $B(t)$, ${t\ge 0}$, and $n$ risky stocks
with prices $S_i(t)$, ${t\ge 0}$, $i=1,2,...,n$, where $n<+\infty$
is given. The prices of the stocks evolve according to the
following equations: \be \label{S}
dS_i(t)=S_{i}(t)\left(a_i(t)dt+\sum_{j=1}^d\s_{ij}(t)
dw_j(t)\right),  \quad t>0, \ee where ($^\top$ denoted transpose)
$w(t)=(w_1(t),\ldots,w_d(t))^\top$ is a standard $d$-dimensional
Brownian motion, $a(t)=(a_{1}(t),\ldots,a_n(t)))^\top$ is the
vector of appreciation  rates, and the $\s_{ij}(t)$ are volatility
coefficients.   The initial prices   $S_i(0)>0$ are given
non-random constants.   The price of the locally riskless asset
evolves according to the following equation \be \label{B}
B(t)=B(0)\exp\left(\int_0^t r(t)dt\right), \ee where $B(0)$ is
taken to be $1$ without loss of generality, and $r(t)$ is the
progressively measurable random interest rate process. Write
$\s(t)=\left\{\s_{ij}(t)\right\}$ for the $n\times d$ dimensional
matrix process. Define the  return to time $t$ by
$dR_i(t)=dS_i(t)/S_i(t),\ R_i(0)=0$, and introduce the vector of
returns
 $R(t)=\left(R_1(t),...,R_{n}(t)\right)^\top$ and of excess returns
 $\ww R_i(t)=R_i(t)-\int_0^t r(\tau)\,d\tau.$
\par
Let $\w r(t)=r(t)(1,...,1)^\top\in{\bf R}^n,\  \ww a(t)=a(t)-\w r(t).$ Then
\be
\label{R}
 d R(t)= a(t)\, dt + \s(t)\, dw(t),\quad
 d\ww R(t)=\ww a(t)\, dt + \s(t)\, dw(t).
\ee
{\bf Remark 2.1.} The first question is how to calibrate this model, i.e. what $a_i$
and $\s_{ij}$ to use. We can observe the prices, hence the returns, interest rate and
excess return, but not the Brownian motion $w$. The volatility  coefficients can in
principle be estimated from $R(\cdot)$ or $\ww R(\cdot)$. In fact,
\be
\label{vol} \int_0^t
\s(\tau)\s(\tau)^\top\,d\tau=<\!R\!\!>_t=<\!\ww R\!\!>_t, \ee the
(observable) quadratic variation process of $R$ or $\ww R$. In
fact we shall {\bf assume} that $\s$ is a non-random function of
the excess return, $\ww R$. It is more difficult to estimate the
appreciation rates $a_i(t)$ in particular because in the short run
the volatility dominates them. We adopt a Bayesian approach.
\par Let
$\{\Fo_t\}$ be the filtration generated by $\{R,r\}$ augmented by
the null sets of ${\cal F}$. It is the observation filtration and
is also generated by $\{S,B\}$ or $\{\ww R,r\}$ (with
augmentation).  To describe the prior  distribution of $a(\cdot)$,
we assume that there exist a separable linear normed space $E$, a
Borel measurable set ${\cal T}\subseteq E$, and a random vector
$\Theta:\O\to{\cal T}$ with the distribution $\nu$. Further  we
assume that we are given measurable functions $A: [0,T]\times
{\cal T}\times C([0,T];{\bf R}^n) \to{\bf R}^n$, $\a: [0,T]\times
C([0,T];{\bf R}^n)\to {\bf R}^{n\times n}$, and $\rho: [0,T]\times
C([0,T];{\bf R}^n)\to {\bf R}$, such that $$ \ww a(t,\o)\equiv
A\left(t,\Theta(\o),\ww R(\cdot,\o)|_{[0,t]}\right),\quad
\s(t,\o)\equiv \a(t,\ww R(\cdot,\o)|_{[0,t]}), \quad r(t,\o)\equiv
\rho\left(t,\Theta(\o),\ww R(\cdot,\o)|_{[0,t]}\right). $$ Here
$f(s,\o)|_{[0,t]}=f(s\wedge t,\o)$.
 Here
$f(s,\o)|_{[0,t]}=f(s\wedge t,\o)$.
\begin{assumption}\label{ass}
$\Theta$ and $w(\cdot)$ are mutually independent;\\
$\sup_{t,f,\theta}(|\rho(t,\theta,f)|+|\a(t,f)|)< \infty\ a.s.$;
there exists a function $K(\cdot)$ and a constant $K_o$ such that
$$\sup_{t,f}|A(t,\theta,f)| \leq K(\theta)<\infty, $$
$$|A(t,\theta,f)-A(t,\theta,g)| \leq
K(\theta)\sup_{\tau\in[0,t]}|f(\tau)-g(\tau)|;$$
$$|\a(t,f)-\a(t,g)|\leq K_o
\sup_{\tau\in[0,t]}|f(\tau)-g(\tau)|;$$
 $\a(t,f)\a(t,f)^\top\geq c I_n,$ where  $c>0$ is a constant and
$I_n$ is the identity matrix in ${\bf R}^{n\times n}.$
\end{assumption}
\par
Let $\O_w \defi C([0,T];\R^n)$, let  $\F_w$ be the completion of
the $\s$-algebra of subsets of $\O_w$ generated by $w(\cdot)$, and
let $\F_{\cal T}$ be the completion of $\sigma$-algebra of subsets
of $\cal T$ generated by $\Theta$. Further, let $\P_w$  be the
probability measure on $\F_w$ generated by $w(\cdot)$.  By the
definitions, $\nu$ is the probability measure on $\F_{\cal T}$.
\par
Without loss of generality, we assume that the probability space
$(\O,\F,\P)$ is such that $\O={\cal T}\times\O_w$,  $\F$ is the
completion of $\F_{\cal T}\otimes\F_w$, and $\P$ is the completion
of $\nu\times\P_w$.
\\
 {\bf Remark 2.2.} (i) The conditions imply
that the solutions of (\ref{S}), (\ref{B}) and (\ref{R}) are
well-defined.\\ (ii) The simplest models have $\ww
a(\cdot)=\Theta(\cdot)$ for a process $\Theta(t)$ independent of
$w(\cdot)$.
 \par As usual it will be productive to work
with an equivalent measure $\P_*$ under which the normalized
wealth process (cf. next section) is a martingale. Set \be
\label{Z} \Z\defi\exp\left(\int_0^T (\s(t)^{-1}\ww a(t))^\top
dw(t)+ \frac{1}{2}\int_0^T |\s(t)^{-1}\ww a(t))|^2dt\right). \ee
Clearly, there exists measurable function $f:{\cal T}\times
\O_w\to\R$ such that $\Z^{-1}=f(\Theta,w(\cdot))$. Since $\Theta$
and $w$ are independent and $|\s(\cdot)^{-1}\ww a(\cdot)|\leq
\sqrt{c}\, K(\Theta)$, then by Fubini's Theorem it follows that $$
\E\Z^{-1}=\int_{\cal
T}\nu(d\theta)\int_{\O_w}P(d\o_w)f(\theta,\o_w)=\int_{\cal
T}\nu(d\theta)1=1, $$ because the Lipschitz conditions on $A$ and
$\a$ allow us to construct $\ww R$, hence $\Z$ for each value
$\theta$
 of $\Theta$.
 Define $\P_*$  by $d\P_*/d\P=\Z^{-1}$.
Let $\E_*$ be the corresponding expectation. Note that
$d\P/d\P_*=\Z$. By Girsanov's Theorem, it follows that the process
$(\ww R(t),\F^{R,r})$ is a martingale with respect to $\P_*$.
\par
 We will require an expression for $\E_*(\Z|\Fo_T)$. To prepare for this,
define $Q(t,\o)\defi (\s(t,\o)\s(t,\o)^\top)^{-1}$,
and for each $\theta\in {\cal T}$, introduce the
process $z(\theta,t)$ as a solution of the equations
\be
\label{za} \left\{\begin{array}{ll} dz(\theta,t)
=z(\theta,t)A\left(t,\theta,\ww R(\cdot)|_{[0,t]}\right)^\top Q(t)\,d\ww R(t),
\\
z(\theta,0)=1.
\end{array}
\right. \ee
Now set
 $$
\oo\Z\defi\int_{{\cal T}}d\nu(\theta)z(\theta,T).
$$
This is the required conditional density of $\P$ with respect to
$\P_*$ given the observations.
\begin{proposition}
\label{prop1}
$\qquad \E_*(\Z|\Fo_T)=\oo \Z.$
\end{proposition}
{\bf Remark 2.3.} Proposition~\ref{prop1} actually holds for more
general $\s$ (not of the form $\a$). It is only required that it
be almost surely pathwise bounded, $\s(t)\s(t)^\top\geq cI_n \
a.s.$, and that $\Theta$, $\s$ and $w$ be mutually independent.
Under these conditions $A$ can also depend additionally on $r$ and
$\s\s^\top$. Now $K(\theta)$ becomes $K(\theta,q,\rho)$ where the
last two arguments stand for the paths of $\s\s^\top$ and $r$
respectively. Proofs are given in the Appendix.

\section{Problem statement}
An investor holds a portfolio of the instruments; the pair $(\pi_0(t), \pi(t))$
describes the portfolio at time $t$:   the process    $\pi_0(t)$ is the  investment in
the bond, $\pi_i(t)$ is the investment in the   $i$th stock,
$\pi(t)=\left(\pi_1(t),\ldots,\pi_{n}(t)\right)^\top$, $t\ge 0$. Let $X_0>0$ be the
wealth of the agent at time $t=0$, i.e. the initial value of the portfolio,  and let
$X(t)$ be the wealth at time $t>0$, $X(0)=X_0$.  Then \be \label{X}
X(t)=\pi_0(t)+\sum_{i=1}^n\pi_i(t). \ee

The portfolio is said to be self-financing if
$   dX(t)=\pi_0(t)\, dr(t)+\pi(t)^\top\,d R(t).
$
For such portfolios
\be\label{XX}
dX(t)=r(t)X(t)\,dt+\pi(t)^\top\, d\ww R(t),
\ee
 \[\pi_0(t)=X(t)- \sum_{i=1}^n\pi_i(t),
\]
so $\pi$ alone suffices to specify the portfolio; it is called a self-financing
strategy. If we define $\ww X(t)\defi B(t)^{-1}X(t)$, then
\be\label{nw}
\ww X(t)=X(0)+\int_0^t B(s)^{-1}\pi(s)^\top \,d\ww R(s).
\ee
For each $\pi$ we denote the corresponding $X$ or $\ww X$ by $X^\pi,\ \ww X^\pi$.
 \par
The investor's problem is to choose $\pi$ according to some criterion. First we note
that the investor must base his decision at time $t$ on his knowledge at time $t$,
which is $\{S(s),r(s):s\leq t\}$ or equivalently $\{R(s),r(s):s\leq t\}$.
 Hence to satisfy the agents observability requirement, $\pi$ must be adapted to $\Fo$.
\par
Let $\F_t$ be the filtration generated by ${R,r,a}$ augmented by
the null sets of $\F$.
\begin{definition}
\label{def1} Let ${\cal A}$ (correspondingly ${\cal A}^a$)
 be the class of all
 $\{\Fo_t\}$-progressively measurable  (correspondingly $\{\F_t\}$-progressively measurable)
processes $\pi(\cdot)$ such that
(i)
$\int_0^{T}|\pi(t)|^2\,dt
<\infty\ \hbox{ a.s.}$ and (ii) there exists a constant
$q_\pi$
such that $\P\{\ww X(t)-X_0\geq q_\pi,\forall t\in[0,T]\}=1$.
\end{definition}
A process $\pi(\cdot)\in {\cal A}$   is said to be an {\em
admissible} strategy. For such $\pi$ the integral in (\ref{nw}) is well defined.
For each $\pi\in{\cal A}$, $\ww  X^\pi(t)$ is a $\P_*$-supermartingale with
$\E_*\ww  X^\pi(t)\leq X_0$ and $\E_* |\ww  X^\pi(t)|\leq |X_0|+ 2|q_\pi|.$
The following definition is standard.

\begin{definition}
\label{repl}
Let $\xi$ be a given random variable.
An admissible  strategy   $\pi(\cdot)$
is said to {\rm replicate} the claim $\xi$  if
$
X^\pi(T)=\xi \ \hbox{ a.s.}
$
\end{definition}
We observe that ${\cal A}^a$ denotes the class of admissible
strategies when no observability requirement is imposed, i.e. the
problem solved first by Merton.
\par
Let $T>0$, let $\w D\subset {\bf R}$ be convex and bounded below, and let $X_0\in \w D$ be given.
Let $U(\cdot):\w D\to{\bf R}\cup\{-\infty\}$ be such that $U(X_0)>-\infty$.
\par
We may state our general problem as follows:
Find   an admissible self-financing strategy $\pi(\cdot)$
which solves the following optimization problem:
\be
\label{c}
\mbox{Maximize}\quad
\E U(\ww X^\pi(T))\quad\hbox{over}\quad
\pi(\cdot)\in\ {\cal A}
\ee
\be
\label{s}
\mbox{subject to }
\left\{
\begin{array}{l}
\ww X^\pi(0)=X_0,
\\
\ww X^\pi(T)\in\w D \quad
\hbox{a.s.}
\end{array}
\right.
\ee
The condition $\ww X^\pi(T)\in\w D$ would represent a requirement
for a minimal normalized terminal wealth if $\w D=[k,+\infty)$, $k>0$.
\par
Roughly speaking, the problem is solved as follows.

Find the optimal terminal value by constrained maximization, then find the optimal
$\pi$ by replicating this terminal value. The extra constraint to ensure possibility
of replication is
$\E_* \ww X(T) = X_0.$
Hence we want to solve:
$\max_\xi\{\E U(\xi)\,\mid\, \xi\in\w D,\ \E_*\xi=X_0\}$, or, using a Lagrange multiplier
$\lambda$ and the fact that $\xi$ is $\Fo_T$ measurable,
\be\label{mmax}
\E_*\max_{\xi\in\w D}\left\{\oo\Z U(\xi)-\lambda\xi\right\} + \lambda X_0.
\ee
To make this program work, we assume that $U$, $X_0$
and $\w D$ satisfy the following three conditions.
\begin{condition}
\label{ass01} There exists a measurable set $\Lambda \subseteq [0,\infty)$,
and a measurable function
$F(\cdot,\cdot):\,(0,\infty)\!\times\Lambda
\to \w D$
such that for each $z>0$,
 $\w x=F(z,\lambda)$ is a solution of
the optimization problem
\be
\label{UU}
\mbox{{\rm Maximize}}\quad
zU(x)-\lambda x
\quad\mbox{{\rm over  }} x\in \w D.
\ee
\end{condition}
This condition allows us to solve the maximization problem in (\ref{mmax}). Of course
the usual concavity hypotheses imply this condition, but more general utility
functions are also covered.
\begin{condition}
\label{assla}
There exists $\w\lambda\in \Lambda$  such that   $\E_* |F(\oo\Z,\w\lambda)|<+\infty$ and
$\E_* F(\oo\Z,\w\lambda)=X_0.$
\end{condition}
With this condition we now know that $\w\lambda$ is the correct multiplier to use. It will also
be seen that the integrability implies that the optimal utility has well defined expectation.
\par
The optimal solution of the problem
(\ref{c})-(\ref{s}) under Conditions \ref{ass01}-\ref{assla}
 was obtained  in the class ${\cal A}^a$
in Dokuchaev and Haussmann (2001) under some additional conditions.
By definition of  ${\cal A}^a$, this solution has the form
$\pi(t)=\G(t,S(\cdot)|_{[0,t]},\ww a(\cdot)|_{[0,t]},r(\cdot)|_{[0,t]})$,
for some measurable function $\G$.
\par
In case $a(\cdot)$ is a Gaussian process and $U=\log$, it is known
that the problem can be solved using Kalman filtering and the
``certainty equivalence" principle, cf Gennotte, (1986). More
precisely, solve the problem as if $a$ were known, to obtain the
optimal strategy as $\pi(t,a)$ and find $m(t)\defi\E\{a(t)|{\cal
F}_t\}$  (the Kalman filter). Then $\pi(t,m)$ is the optimal
solution of the given problem. This result is incorrect for
non-$\log$ utility functions, cf Kuwana, (1995); however we can
resurrect it for some other utility functions if we allow proxies
for $a$ other than $m$.
\begin{definition}
\label{defest}
Let $\pi(t)=\G(t,S(\cdot)|_{[0,t]},\ww
a(\cdot)|_{[0,t]},r(\cdot)|_{[0,t]})$ be an optimal solution of the
problem (\ref{c})-(\ref{s}) in the class ${\cal A}^a$, where
$\G$ is a measurable function.  Further, let
$\w\pi(t)$ be an optimal solution of the problem (\ref{c})-(\ref{s})
in the class ${\cal A}$, and let there exists a $n$-dimensional
$\{\Fo_t\}$-adapted random vector  process  $\w a(t)$ such that
$\w\pi(t)\equiv
\G(t,S(\cdot)|_{[0,t]},\w a(\cdot)|_{[0,t]},r(\cdot)|_{[0,t]})$.
Then $\w a(t)$ is said to be the {\rm equivalence filter} of  $\ww a(t)$
with respect to the problem (\ref{c})-(\ref{s}).
\end{definition}
Note that we {\it do not} assume that $\w a(t)$ is a function of
the current conditional distribution ${\cal P}_{\ww
a(t)}(\cdot\,|\,\Fo_t)$ of $\ww a(t)$.
\section{Existence of the optimal claim and strategy}
\label{sec4}
 We solve our problem in two steps. First we show that $\E U(F(\oo
\Z,\w\lambda))$ is an upper bound for the expected utility of normalized terminal
wealth for $\pi(\cdot)\in{\cal A}$. Then we show that a portfolio $\w\pi(\cdot)$ which
replicates the claim $B(T)F(\oo \Z,\w\lambda)$ exists. This establishes the optimality
of $\w\pi(\cdot)$. We exhibit $\w\pi$ for a couple of utility functions in the next
section, and then treat the general case in the following one.
\par
Let $U^+(x)\defi \max(0,U(x))$, $U^-(x)\defi\max(0,-U(x))$.
Let $F(\cdot)$ be as in Condition \ref{ass01}.
\begin{theorem}
\label{ThM1} Under Assumption~\ref{ass} and Conditions~\ref{ass01}, \ref{assla},  let
\be
\label{wxi}
\w\xi\defi F(\oo\Z,\w\lambda)
\ee
with $\w\lambda$ as in  Condition \ref{assla}. Then
\par
{\rm (i)} $\E\, U^-(\w\xi)<\infty$,  $\w\xi\in \w D$ a.s.
\par
{\rm (ii)} $\E U(\w\xi)\ge \E U(\ww X^\pi(T))$,
$\forall \pi(\cdot)\in{\cal A}$.
\par
{\rm (iii)}
The claim $B(T)\w\xi$ is attainable in ${\cal A}$, and
there exists a  replicating strategy in  ${\cal A}$.
This strategy is optimal for the problem  (\ref{c})-(\ref{s}).
\end{theorem}
The proof is in the Appendix.
\par
{\bf Remark 4.1} It now follows that the optimal terminal wealth is
$B(T)F(\oo\Z,\w\lambda)$ and the optimal strategy is determined implicitly by
replication. So we have a type of equivalence principle: proceed as for the completely
observable problem, but replace $\Z$, the density of $\P$ with respect to $\P_*$, by
$\oo\Z$, the conditional expectation of $\Z$, cf Proposition~\ref{prop1}.
\par
The first two parts of the theorem hold under the weaker conditions mentioned in
Remark 2.3, but then we cannot appeal to the martingale representation theorem to
obtain the replication of part (iii). If we have another technique for establishing
this replication, then the theorem holds under the weaker hypotheses. We pursue this idea in the next section.
\section{Replication with myopic strategies and equivalence filters}
\label{sec4'} We now consider two special utility functions,
$U(x)=\log (x+\d)$, $\d\geq 0$, and $U(x)=x^\d$ for some $\d$, but
under the weaker assumptions of Remark 2.3. In these cases we can
compute the replicating strategy directly and so solve the problem
explicitly. We also find equivalence filters of the excess
accumulation rates $\ww a_i$.
\begin{lemma}
\label{Thlog}
Let $U(x)\equiv \log (x +\delta)$, $\delta\geq 0$, $X_0>0$ and
$(0,+\infty)\subseteq \w D$.
Then the optimal solution in the class
${\cal A}$ of the problem (\ref{c})-(\ref{s}) is
\be
\label{logpi}
\begin{array}{rcl}
\w\pi(t)^\top&\defi & (X_0+\delta)B(t)\int_{{\cal T}}d\nu(\theta)
 z(\theta,t)
 A\left(t,\theta,\ww R(\cdot)|_{[0,t]},\s(\cdot)\s(\cdot)^\top|_{[0,t]},r(\cdot)|_{[0,t]}
 \right)^\top Q(t)\\
&=&(X^{\w\pi}(t)+\delta B(t))\frac{\int_{{\cal T}}d\nu(\theta)z(\theta,t)
A\left(t,\theta,\ww
R(\cdot)|_{[0,t]},\s(\cdot)\s(\cdot)^\top|_{[0,t]},r(\cdot)|_{[0,t]} \right)^\top
}{\int_{{\cal T}}d\nu(\theta)z(t,\theta)}Q(t)
\end{array}
\ee
and
\be
\label{logX}
X^{\w\pi}(t)=B(t)\left((X_0+\delta)\int_{{\cal T}}d\nu(\theta)
 z(\theta,t)-\delta\right)\ \mbox{for all }t.
\ee
\end{lemma}
\par
{\em Proof:}
We must replicate the claim $B(T)\w\xi$. According to Condition~\ref{ass01},
$F(z,\lambda)=z/\lambda-\delta$, so Condition~\ref{assla} gives
$\w\lambda=\E_*\oo\Z/(X_0+\delta)=1/(X_0+\delta)$ since
$$
\E_*\oo\Z=\E_*\E_*(\Z|\Fo_T)=\E_*\Z=\E\Z^{-1}\Z=1.
 $$
Write $X_\delta$ for $X_0+\delta$. It follows that
\be \label{eq23}
\begin{array}{l}\w\xi=F(\oo\Z,\w\lambda)=X_\delta\oo\Z-\delta \\
\quad = X_\delta\left\{\int_{{\cal T}}d\nu(\theta)\left[1+\int_0^T
z(\theta,t)A\left(t,\theta,\ww R(\cdot)|_{[0,t]},\s(\cdot)\s(\cdot)^\top|_{[0,t]},r(\cdot)
|_{[0,t]}\right)^\top Q(t)d\ww R(t)\right]\right\}-\delta\\
\quad = X_0 + \int_0^T B(t)^{-1}\w\pi(t)^\top\,d\ww R(t) = \ww X(T)
\end{array}
\ee
if
$$\w\pi(t)^\top=B(t)X_\delta \int_{{\cal T}}d\nu(\theta)z(\theta,t)
A\left(t,\theta,\ww R(\cdot)|_{[0,t]},\s(\cdot)\s(\cdot)^\top
|_{[0,t]},r(\cdot)|_{[0,t]}\right)^\top Q(t). $$
 Hence this strategy replicates
$B(T)\w\xi$ and so is optimal.
\par
 Moreover
$$X^{\w\pi}(t)=B(t)\ww X^{\w\pi}(t)= B(t)
\left(X_0 + \int_0^t B(s)^{-1}{\w\pi}(s)^\top\,d\ww R(s)\right)=B(t)\left(X_\delta
\int_{{\cal T}}d\nu(\theta)z(t,\theta)-\delta\right),$$
 so in fact
$$\w\pi(t)^\top=(X(t)+\delta B(t))\frac{\int_{{\cal T}}d\nu(\theta)z(\theta,t)
A\left(t,\theta,\ww R(\cdot)|_{[0,t]},\s(\cdot)\s(\cdot)^\top|_{[0,t]},r(\cdot)|_{[0,t]}
\right)^\top }{\int_{{\cal T}}d\nu(\theta)z(t,\theta)}Q(t).$$
$\Box$
\begin{corollary}
\label{corlog} {\rm (i)} Under the conditions of Lemma
\ref{Thlog}, the equivalence filter of $\ww a(t)$ is \be
\label{est} \w a(t)=\frac{\int_{{\cal T}}d\nu(\theta)
 z(\theta,t)A\left(t,\theta,\ww R(\cdot)|_{[0,t]},\s(\cdot)\s(\cdot)^\top|_{[0,t]},r(\cdot)|_{[0,t]}\right)}{\int_{{\cal T}}d\nu(\theta)
 z(\theta,t)}.
\ee
\par
{\rm (ii)} Assume that $\E |K(\Theta,\s(\cdot),\s(\cdot)^\top,r(\cdot))|^2<\infty$.
The process $\w a(t)$ defined by (\ref{est}) is such that
$\w a(t)=\E\{\ww a(t)|\Fo_t\}$, i.e. it is
the minimum variance estimate
in the class of estimates
based on observations of $(S,r)$ (or $(R,r)$) up to
time $t$   assuming the prior $\nu$ for $\Theta$. The optimal expected utility   is
\be
\label{EU}
\E \log (\ww X^{\w\pi}(T)+\delta)
=\frac{1}{2}\E\int_0^T\w a(t)^\top Q(t)
\w a(t) dt+\log (X_0+\delta).
\ee
\end{corollary}
\par
Part (i) is obvious if we recall that the optimal strategy in ${\cal A}^a$ is
$\pi(t)^\top=(X(t)+\delta B(t))\ww a(t)^\top Q(t)$, cf. Dokuchaev and Haussmann (2001)
(or assume that ${\cal T}$ is a singleton), and use (\ref{logpi}). We give the proof
of part (ii) in the Appendix. Observe that if $\ww a(t)$ is conditionally Gaussian,
i.e.
$$d\ww a(t)=\left(c_1(t,\ww R(\cdot)|_{[0,t]},r(\cdot)|_{[0,t]})
-c_2(t,\ww R(\cdot)|_{[0,t]},r(\cdot)|_{[0,t]})\ww a(t)\right)dt
+ c_3(t,\ww R(\cdot)|_{[0,t]},r(\cdot)|_{[0,t]})dw'$$
with $\Theta=w'(\cdot)$ an independent Brownian motion,
then the Kalman filter can be used to calculate $\w a$
and hence $\w\pi(t)=X^{\w\pi}(t)Q(t)\w a(t)$. This result extends Example 4.4 of Lakner (1998).
\par
We can now characterize $\oo\Z$ further; this will be helpful in Section~ \ref{sec6}.
We add that the following corollary also delivers the result of Lakner (1998),
Theorem~3.1, under our more general assumptions.
\begin{corollary}\label{cor23}
Define $\oo\Z(t)\defi\E_*(\oo\Z|\Fo_t)$, $\w a(t)\defi \E(\ww a(t)|\Fo_t)$. Then $\oo\Z=\oo\Z(T)$ and
\be \label{deZ}
\oo\Z(t)=\exp{\left\{\int_0^t\w a(s)^\top Q(s)\,d\ww R(s)-\frac{1}{2}\int_0^t\w a(s)^\top Q(s)\w a(s)\,ds\right\}}.
\ee
\end{corollary}
{\em Proof:}
 We take $\d=0$ and $X_0=1$.
Then (\ref{eq23}) implies that $\oo\Z=\ww X^{\w\pi}(T)$, so
$\oo\Z(t)=\ww X^{\w\pi}(t)$ since the latter is a $(\P_*,\F^{R,r
}_t)$-martingale; hence $\log\oo\Z(t)=Y(t,\w\pi)$ where $Y$ is as
in the proof of Corollary~\ref{corlog}. The result follows from
(\ref{YY}). $\Box$
\par
We can also carry out this program for certain power utility functions, those for
 which the function $F$ has the form $F(z,\lambda)= Cz^l$ with $l>1$  an integer, i.e.
 $U(x)=x^\delta /\delta$ with $\delta={1-\frac{1}{l}}$,  if we
 make further assumptions on the $A$ and $\s$. Specifically, $A$ should be linear in
 $\theta$ and $\int_0^T |\s^{-1}A(t,\theta,\ww R,\s\s^\top,r)|^2dt$ must be
 deterministic, which realistically means that we take $\Theta$ to be a process, $\ww
 a(t)=\Theta(t)$, and $\s$ is non-random.
 Moreover we need some integrability, i.e.
\be \label{GG}
G\defi \int_{{\cal T}^{l}} d\nu(\theta_1)\cdots
d\nu(\theta_{l})\g(\theta_1,...,\theta_l)<\infty,
\ee
where
$$
\g(\theta_1,...,\theta_l)\defi \exp\left\{
\sum_{\stackrel{i,j=1}{i<j}}^{l} \int_0^T
\theta_i(t)^\top
Q(t)\theta_j(t)\,dt\right\}. $$
Here each $\theta_i$ is an $n$-dimensional function, a sample path of $\ww a$.

We note that $\g$ and $G$ are non-random. It is convenient to introduce the notation
$\oo\T\defi
\{
\sum_{k=1}^l\theta_k\,:\,\theta_i\in{\cal T},
\ i=1,\ldots,l \}$,
let $\chi_{_D}$ be the indicator of $D$ and
define a measure $\oo\nu$ on $\oo\T$ by
\[\oo\nu(D)\defi
\frac{\int_{\T^l}\chi_{_D}(\sum_1^l\theta_k)\,d\nu(\theta_1)\cdots d\nu(\theta_l)
\g(\theta_1,...,\theta_l) }{\int_{\T^l}d\nu(\theta_1)\cdots d\nu(\theta_{l})
\g(\theta_1,...,\theta_{l})
}.\]
\par
\begin{theorem}
\label{Thm}  Assume that $A(\cdot,\theta,f,q,\rho)=\theta(\cdot)$, $\s$  deterministic,
 $(0,+\infty)\subseteq \w D$,  $X_0>0$, $U(x)\equiv x^\d/\d$,
$\d={(l-1)/l}$ for some integer $l>1$, and $G<\infty$. Then
\par
{\rm (i)}   ${F}(z,\lambda)\equiv z^l \lambda^{-l}$ and
$\w\lambda= X_0^{-1/l}(\E_*\oo\Z^l)^{1/l}$.
\par
{\rm (ii)}
The optimal solution in the class
${\cal A}$ of the problem (\ref{c})-(\ref{s}) is
\be
\label{powpi}
\begin{array}{lll}
\w\pi(t)^\top&\defi& {X_0}B(t)\int_{\oo\T}d\oo\nu(\theta)
 z\left(\theta,t\right)\theta(t)^\top Q(t)\\
&=&X^{\w\pi}(t)\frac{\int_{\oo\T}d\oo\nu(\theta)
 z\left(\theta,t\right)\theta(t)^\top}{\int_{\oo\T}d\oo\nu(\theta)
 z\left(\theta,t\right)} Q(t),
\end{array}
\ee
and
\be
\label{powX}
X^{\w\pi}(t)={X_0}B(t)\int_{\oo\T}d\oo\nu(\theta)
 z\left(\theta,t\right).
\ee
Moreover
\be
\label{Ezm=2}
\E U(\ww X^{\w\pi}(T))
=X_0^{\d} G^{1-\d}/\d.
\ee
\end{theorem}
\par
{\it Remark}. We will see below  that the
equivalence filter $\w a(t)$ under assumptions of Theorem \ref{Thm}
{\it differs} from  $\E\{\ww a(t)|\Fo_t\}$
and, in general, is not a function
of the current conditional
distribution ${\cal P}_{\ww a(t)}(\cdot\,|\Fo_t)$ of $\ww a(t)$.
However, we can write $\w a$ as a conditional expectation of $\ww a$ if we change measure.
Let  $\oo\P$ be given by $\P$ when we replace $\nu$ defined on ${\cal T}$ by $\oo\nu$ defined on  ${\oo\T}$.
\par
\begin{corollary}
\label{El} Under the conditions of Theorem \ref{Thm}, the
equivalence filter of $\ww a(t)$  is \be \label{estm} \w
a(t)=\frac{Q(t)^{-1}\w\pi(t)}{lX^{\w\pi}(t)}=l^{-1}\oo\E \{\ww
a(t)|\Fo_t\} \ee where $X^{\w\pi}(t)$ is the wealth defined by
(\ref{powX}).
\end{corollary}
\par
In particular, if $\ww a(\cdot)=\Theta$ is time independent, Gaussian, with density function
$\varphi$, then $G<\infty$ only if $\int_0^T Q(t)\,dt$
is so small that $\log\Phi(x_1,\ldots,x_l)$ is a negative
definite quadratic form (plus an affine term), where
$$
\Phi(x_1,\ldots,x_l)\defi\left\{\varphi(x_1)\cdots \varphi(x_l)
\exp\sum_{\stackrel{i,j=1}{i<j}}^l x_i^\top \left(\int_0^T Q(t)\,dt\right)  x_j\right\}.
$$
It follows that
if $\ww\Theta=(\Theta_1,\ldots,\Theta_l)$ is defined to have density function
$\Phi(x_1,\ldots,x_l)/G$,
then $\ww\Theta$ is Gaussian, hence the distribution of $\sum_{i=1}^l \Theta_i$,
which is $\oo\nu(\cdot)$, is Gaussian.
 This means that Kalman filtering can be employed to calculate
$\w a(t)$. \par Cvitani\'c {\it et all} (2002) presented the
explicit optimal strategy for non-observable parameters for the
Gaussian prior for $U(x)=\d^{-1}x^\d$, but only for the case when
$\d<0$.  Our approach is quite different and covers $\d>0$ but
only for $\d=(l-1)/l$, $l=2,3,...$. \par The two special cases
discussed above are of limited interest because of the special
nature of the utility functions used even though we have
generalized the market dynamics somewhat ($r$ random in both
cases and $\s$ random in the first). Let us then find the optimal
strategies for more general utility functions but under our more
restrictive assumptions, cf. Assumption~\ref{ass}.
\section{Embedding to a Markovian setting for the general utility}
\label{sec6} We can use a PDE-based approach to replication,
hence to the solution of our problem, if the claim to be
replicated, here a function of $\oo\Z$, is a functional of a
Markov process. We mention three examples below. Suppose there
exist an  integer $M>0$, a deterministic function $\phi:\R^M\to
\R$, and a $M$-dimensional Markov process $y(\cdot)$  such that
$$ \oo\Z=\phi(y(T)), $$ and $y(\cdot)$ is the solution of an
It\^o equation \be \label{yy}
 \left\{
\begin{array}{ll}
dy(t)=f(y(t),t)dt+b(y(t),t)\,d\ww R(t),\\
y(0)=y_0\in\R^M,\end{array}\right.
\ee
 where
$f(\cdot):\R^M\times\R\to\R^M$, $b(\cdot):\R^M\times\R\to\R^{M\times n}$ are
measurable functions. We can always append the equation $d\ww R=d \ww R$ so we may
assume that $\ww R$ is included in $y$ if needed. Then we assume that $\a(t,\ww R(\cdot))=\a(t,
y(t))$. Write $\oo b(y,t)$ for $b(y,t)\a(t,y)$.
\par
 We assume that  the functions
$\oo b(y,t)$, $f(y,t)$ are H\"older and such that
$$
|\oo b(y,t)|+|f(y,t)|\le \const(|y|+1).
$$
Further, we assume that $\p
\oo b(y,t)/\p y$, $\p^2 \oo b(y,t)/\p y^2$, $\p f(y,t)/\p y$ and $\p^2
f(y,t)/\p y^2$ are uniformly bounded and  H\"older.
\par
Let $y_*(\cdot)$ denote the solution of (\ref{yy}) with $\ww R(\cdot)$ replaced by
$\ww R_*(\cdot)=\int_0^\cdot\a(t,\ww R_*(t))\,dw(t)$ and introduce the Banach space
$\Y^1$ of functions
$u:\,\R^M\times[0,T] \to\R$ with the norm $$
\|u(\cdot)\|_{\Y^1}\defi\left( \sup_t\E |u(y_*(t),t)|^2+\E
\int_0^T\left|\frac{\p u}{\p x}(y_*(t),t)\right|^2dt\right)^{1/2}.
$$
\begin{proposition}
\label{prop01} Let $C(\cdot):\,\R^M\to\R$ be a measurable function such that $\E
C(y_*(T))^2<+\infty$ and  $\E C(y_*(T))=X_0$. Then there exists an admissible strategy
$\pi(t)=( \pi_1(t),\ldots,\pi_{n}(t))\in {\cal A}$ which replicates the claim
$B(T)C(y(T))$. Furthermore,
$$
 \pi(t)=B(t)b(y(t),t)^\top \frac{\p V}{\p y}(y(t),t),\quad
\ww X^\pi(t)=V(y(t),t),
$$
where $\frac{\p V}{\p y}$ denotes the gradient of $V$ with respect to it's first
argument and the function $V=V(y,t):\,\R^M\times [0,T]\to \R$ is such that
\begin{eqnarray}\label{parab1}
 \frac{\p V}{\p t}(y,t)+\frac{\p V}{\p y}^\top(y,t) f(y,t)+
\frac{1}{2}{\rm Tr}\{\frac{\p^2 V}{\p y^2}(y,t)\, \oo b(y,t)\oo b^\top(y,t)\}=0,\\
 \label{parab1'}
V(y,T)=C(y).\end{eqnarray} The problem
(\ref{parab1})--(\ref{parab1'}) admits a solution in the class
$\Y^1$.
\end{proposition}
 Let $V(x,t,\lambda):\,\R^M\times[0,T]\times\Lambda\to \R$ be
the solution of the partial differential equation (\ref{parab1}) with the condition
 \be
 \label{parab2}
 V(y,T,\lambda)=F(\phi(y),\lambda).
 \ee The following result now is immediate.
 \begin{theorem}\label{ThMarkov} Let the function $F(\cdot)$ be such that
 \be\label{q2}
 \E_* F(\oo\Z,\w\lambda)^2<+\infty.
 \ee
With $\w\lambda$ as in  Condition \ref{assla}, there exists an
admissible self-financing strategy $\pi(\cdot)\in {\cal A}$ which
replicates the claim $B(T)F(\oo\Z,\w\lambda)$. This
strategy is an optimal solution of the problem
(\ref{c})-(\ref{s}), and
\be
\label{Xopt}
\pi(t)=B(t)b(y(t),t)^\top \frac{\p V}{\p
y}(y(t),t,\w\lambda),\quad
\ww X^\pi(t)=V(y(t),t,\w\lambda).
 \ee
\end{theorem}
\par
{\bf Example 1.} Let us repeat briefly the solution from Dokuchaev (2005) for the problem  solved first in Lakner (1998) by a different method. 
Both solutions  involved the Kalman
filter. Assume that we are given measurable deterministic processes $\a(t)$, $\b(t)$,
$b(t)$ and $\d(t)$ such that \be\label{a-eq} d\ww a(t)=\a(t)[\d(t)-\ww
a(t)]dt+b(t)d\ww R(t)+\b(t) dW(t), \ee where $\a(t)\in\R^{n\times n}$,
$\b(t)\in\R^{n\times n}$, $b(t)\in\R^{n\times n}$, $\d(t)\in\R^n$, and where $W$ is an
$n$-dimensional Wiener process in $(\O,\F,P)$, independent on $w$ under $\P$. We
assume that $\a(t)$, $\b(t)$, $b(t)$, and $\d(t)$ are H\"older in $t$ and such that
the matrix $\b(t)$ is invertible and $|\b(t)^{-1}|\le c$, where $c>0$ is a constant.
Further, we assume that $\ww a(0)$ follows an $n$-dimensional normal distribution with
known mean vector $m_0$ and covariance matrix $\g_0$.  We note that this setting
covers the case when $\ww a$ is an $n$-dimensional Ornstein-Uhlenbeck process with
mean-reverting drift.
\par
Let $y(t)=(y_1(t),...,y_{n+2}(t))=(\w y(t),y_{n+1}(t),y_{n+2}(t))$ be a process in
$\R^{n+2}$, where
$$ \ba  \w y(t)=\E\{\ww a(t)|\Fo_t\},\\
y_{n+1}(t)=\int_0^t\w y(s)^\top Q(s)\,d\ww R(s),\\
y_{n+2}(t)=\exp\left(-\frac{1}{2}\int_0^t\w y(s)^\top Q\w y(s)ds\right).\ea
$$
\par
Clearly, $y_{n+2}(t)\in (0,1]$, thus, $\psi(y_{n+2}(t))\equiv
y_{n+2}(t)$. Theorem 10.3 from Liptser and Shiryaev (2000), p.396,
gives the equation for $\w a(t)=\w y(t)$ such that the equation
for $y(t)$ is
$$ \ba d\w y(t)=[A(t)\w y(t)+\a(t)\d(t)]dt + [b(t)\s(t)^\top+\g(t)]Q(t)\,d\ww R(t),
\\
dy_{n+1}(t)=\w y(t)^\top Q(t)\,d\ww R(t),
\\
dy_{n+2}(t)=-\frac{1}{2}\psi(y_{n+2}(t))\w y(t)^\top Q(t)\w y(t)dt. \ea
$$
Here $\g(t)$ is $n\times n$-dimensional matrice defined from the Riccati's equation
\be\label{gamma}
\left\{\ba\frac{d\g}{dt}(t)=-[b(t)\s(t)^\top+\g(t)]Q(t)[b(t)\s(t)^\top+\g(t)]^\top-\ww\a(t)\g(t)
-\g(t)\ww\a(t)^\top+\b(t)\b(t)^\top,\\
\g(0)=\g_0,
 \ea\right.\ee
  $ A(t)\defi -\ww\a(t)-\g(t)Q(t)$. Note that the  the corresponding $f,b$
satisfy the required conditions. Since $\s$ is independent of $\ww R$ then $\ww R$ is
not required as a component of $y$.
\par
Therefore, the equation for $y(t)$ can be written as (\ref{yy}) and the corresponding
$f,b$ satisfy the required conditions. Since $\s$ is independent of $\ww R$ then $\ww
R$ is not required as a component of $y$.
\par
By Corollary \ref{cor23}, it follows that $\oo\Z=\phi(y(T))$, where the function
$\phi(\cdot):\R^{n+2}\to\R$ is such that $\phi(y)=y_{n+2}\exp{y_{n+1}}$ for
$y=(y_1,\ldots,y_{n+1},y_{n+2})$. Thus, all assumptions of Theorem \ref{ThMarkov} are
satisfied if (\ref{q2}) is satisfied. In particular, if $F$ is  bounded then
(\ref{q2}) is satisfied; if $F(\oo Z,\b,\lambda)$ is polynomial with respect to
$\oo\Z$, then (\ref{q2}) is satisfied if the variance of $\ww a(t)$ is small enough.
\par
 Note that the solution in Lakner
(1998) express the optimal strategy via a conditional expectation
of an optimal claim; our solution borrowed from Dokuchaev (2005)  is more constructive provided we
can solve the Cauchy problem (\ref{parab1}), (\ref{parab2}).
\par
For an Euclidean space $E$ we shall denote by $B([0,T];E)$  the
set of bounded measurable functions $f(t):[0,T]\to E$.
\par {\bf Example 2.} Assume that the number of
possible paths of $\ww a$ is finite. Assume  that $\s$ is
non-random, $A(t,\theta,f)=\theta(t)$  and there exist an integer
$d>1$ and a set $\{\theta_i(\cdot): i = 1,\ldots, d\}\subset
B([0,T];{\bf R}^n)$ such that $\sum_{i=1}^d p_i=1$ where $p_i\defi
\P(\ww a(\cdot)=\theta_i(\cdot))$. Set $y(t)\defi
(y_1(t),...,y_d(t))^\top$, where $y_i(t)\defi z(\theta_i,t)$. Let
$\w F(y)\defi F(p^\top y,\w \lambda)$, $D\defi
(0,+\infty)^d\times[0,T)$ and let  $b(y,t):D\to {\bf R}^{d\times
n}$, be such that the $i$th row of $b$ is $y_i\theta_i(t)^\top$.
Then $dy(t)=b(y(t),t)Q(t)\,d\ww R$ and
(\ref{parab1}),(\ref{parab2}) becomes $$\left\{
\begin{array}{ll}
\frac{\p V}{\p t}(y,t)+\frac{1}{2}
Tr[\frac{\p^2 V}{\p y^2}(y,t)b(y,t)Q(t) b(y,t)^\top]=0,\\
V(y,t)\to \w F(y) \quad \hbox{as}\quad t\to T-0 \quad \forall y.
\end{array}
\right.
$$
Note that the equation  is degenerate in general, so it may not be easy to solve.
Nevertheless, the theorem gives the optimal strategy in terms of $V$.
\par
{\bf Example 3.} Here  $\ww a(t)$ evolves as a function of a finitely-valued
Markov process. For simplicity, let $n=1$. Assume that $\ww a(t)=A(t,\theta(t),R(t))$
and $\s(t)=\a(t,R(t))$, where  the process $\t(t)$ is a random Markov process such
that $\P(\t(t)\in \Lambda)=1$, and $\Lambda=\{\t_i: i = 1,\ldots, d\}$ is a given
finite set, $d>1$ is a integer. We assume that  $A(t,\cdot):\Lambda\times
\R\to\R$ and $\a(t,\cdot):\R\to\R$ are given measurable functions satisfying
Assumption~\ref{ass}. We are given the initial distribution of $\t(0)$, i.e.
we are given $\oo
y_i\defi P(\theta(0)=\t_i)$, and we are given bounded functions
$l_{ij}(\cdot):[0,T]\to\R$ such that
$$
p_{ij}(t,s)=\d_{ij}+\int_s^t\sum_{k=1}^d l_{ki}(\tau)p_{kj}(\tau,s)d\tau \quad \forall
s\le t,
$$
where  $p_{ij}(t,s)\defi P(\t(t)=\t_i|\t(s)=\t_j)$, and where $\d_{ij}$ is the
Kronecker delta, cf. Liptser and Shiryaev (2001), Lemma~9.1. This specifies $\nu$.
\par
 Set $M\defi d+2$, $y(t)\defi
(y_1(t),...,y_M(t))^\top$, where $$ \ba y_i(t)\defi P(\theta(t)=\t_i\,|\,\Fo_t),
\quad
i=1,\ldots d,\\
y_{d+1}(t)\defi \ww R(t);\qquad y_{d+2}(t)\defi \oo\Z(t). \ea
$$
By Theorem 9.1 from Liptser and Shiryaev (2001), p.355, we have
\be\label{fm} \left\{\ba &dy_i(t)= \sum_{k=1}^d
l_{ki}(t)y_{k}(t)dt +{y_i(t)}{\a(t,\ww R(t))^{-2}}\biggl[
A(t,\t_i,\ww R(t))\\&\hphantom{xxxxxx}-\sum_{k=1}^d A(t,\t_{k},\ww
R(t))y_{k}(t)\biggr]\biggl[d\ww R(t)
-\sum_{k=1}^d A(t,\t_{k},\ww R(t))y_{k}(t)\,dt\biggr], \\
&y_i(0)=\oo y_i,\qquad i=1,\ldots d.\ea\right.
 \ee
\par
 To keep a linear bound we introduce a bounded smooth function  $\psi(\cdot)\in C^{\infty}(\R)$ such that
 $\psi(x)=x$ ($\forall x\in[0,1]$) (clearly, there exists such a function). As $y_i\in[0,1], i=1,\ldots,d$, we may replace such $y_i$ by $\psi(y_i)$ as needed. Then we have
$$
 \w a(t)\defi \E\{\ww a(t)|\Fo_t\}= \sum_{i=1}^d
A(t,\t_i,y_{d+1}(t))y_i(t)=\sum_{i=1}^dA(t,\t_i,y_{d+1}(t)) \psi(y_i(t)),
$$
so again (\ref{deZ}) and (\ref{fm}) imply \be\label{fm1}
\left\{\ba dy_i(t)= \sum_{k=1}^d l_{ki}(t)y_{k}(t)dt
+{y_i(t)}{\a(t,\ww R(t))^{-2}}\biggl[
A(t,\t_i,\ww R(t))\\
\hphantom{xxxxxx}-\sum_{k=1}^d
A(t,\t_{k},\ww R(t))\psi(y_{k}(t))\biggr]\biggl[d\ww R(t)
-\sum_{k=1}^d A(t,\t_{k},\ww R(t))\psi(y_{k}(t))\,dt\biggr], \\
dy_{d+1}(t)= d\ww R(t),\\
dy_{d+2}(t)=y_{d+2}(t)\sum_{i=1}^d A(t,\t_i,y_{d+1}(t))\psi(y_i(t))\a(t,y_{d+1}(t))^{-2}d\ww R(t), \\
y_i(0)=\oo y_i,\ i=1,\ldots d,\ y_{d+1}(0)=0,\  y_{d+2}(0)=1.\ea\right.
 \ee
Clearly, the system of equations (\ref{fm})-(\ref{fm1}) can be
rewritten in the form of (\ref{yy}), the corresponding $f,\bar b$
satisfy the required conditions, and  all assumptions of Theorem
\ref{ThMarkov} are satisfied, and the optimal strategy can be
found from the corresponding equation (\ref{parab1}),
(\ref{parab2}).
\section{Appendix: Proofs}
\label{sec8}
 First we prove Proposition~\ref{prop1}. To this end define
$$\ww R_*(t)\defi \int_0^t \a(\tau,\ww R_*(\cdot))\, dw(\tau),\qquad \ww a_*(t)\defi
A(t,\Theta,\ww R_*(\cdot)|_{[0,t]}),$$
\be
\label{Z*} \Z_*\defi\exp\left(\int_0^T (\a(t,\ww R_*(\cdot))^{-1}\ww a_*(t))^\top
dw(t) -\frac{1}{2}\int_0^T |\a(t,\ww R_*(\cdot))^{-1}\ww a_*(t))|^2dt\right). \ee
\begin{proposition}
\label{psi1} There exists a measurable function
$\psi:C([0,T];{\bf R}^n)\times B([0,T];{\bf R}^n)\to{\bf R}$ such that $\Z_*=\psi(\ww R_*(\cdot),\ww
a_*(\cdot))$ and $\Z=\psi(\ww R(\cdot),\ww
a(\cdot))$ a.s. Moreover,
$z(\theta,T)=\psi(\ww R(\cdot),A(\cdot,\theta,\ww R))$.
\end{proposition}
\par
{\it Proof}.
Define
\be \label{V=V}
{\cal Q}(t,f)\defi \a(t,f)\a(t,f)^\top.
\ee
 Then
\be
\label{lZ}
 \log\Z=
\int_0^T \ww a(t)^\top{\cal Q}(t,\ww R(\cdot)|_{[0,t]})\left(
d\ww R(t)-\frac{1}{2}\ww a(t)dt\right),
\ee
and
$$
\log\Z_*= \int_0^T \ww a_*(t)^\top{\cal
Q}(t,\ww R_*(\cdot)|_{[0,t]})\left(  d \ww R_*(t)-\frac{1}{2}\ww a_*(t)dt\right).
$$
This defines $\psi$.
\par
Since $z(t,\theta)$ satisfies (\ref{lZ}) with $\ww a(\cdot)$ replaced by $A(\cdot,\theta,\ww R)$, the last result follows. \\
$\Box$
\par
Let
\be
\label{lZb}
 \oo\Z_*\defi \int_{\cal T}
d\nu(\theta)\psi(\ww R_*(\cdot), A(\cdot,\theta,\ww
R_*(\cdot)))\defi\oo\psi(\ww R_*(\cdot)). \ee It follows from
Proposition~\ref{psi1} that $\oo\Z=\oo\psi(\ww R(\cdot))$.
Finally, since $\Theta$ is independent of $w,r$, hence of $\ww
R_*,r$, it follows that \be\label{eZ} \oo \Z_* = \E(\Z_* | \ww
R_*,r). \ee
\begin{proposition}
\label{propSS} Let $\phi:C([0,T];{\bf R}^n)\times B([0,T];{\bf R}^n)\times B([0,T];{\bf R})\to{\bf R}$ be a function such that
$\E\phi^-(\ww R(\cdot),\ww a(\cdot),r(\cdot))<+\infty$ and let
$\w\phi$ be a similar function but with no dependence on $\ww a$.
Then
\be
\label{phiphi2} \E\phi(\ww R(\cdot),\ww a(\cdot),r(\cdot))
=\E\Z_*\phi(\ww R_*(\cdot),\ww a_*(\cdot),r(\cdot)),
\ee
\be
\label{phiphi3} \E\w\phi(\ww R(\cdot),r(\cdot))
=\E\oo\Z_*\w\phi(\ww R_*(\cdot),r(\cdot)),
\ee
\be
\label{phiphi4} \E_*\w\phi(\ww R(\cdot),r(\cdot))
=\E\w\phi(\ww R_*(\cdot),r(\cdot)).
\ee
\end{proposition}
\par
{\it Proof}.  By assumption $\Theta$ is independent of $w(\cdot)$. Define the probability measure $\w\P$ by $d\w\P/d\P=\Z_*$.
Then $\E\{\Z_*|\Theta\}=1$ and
 to prove  (\ref{phiphi2}) it suffices
to prove
\be
\label{cond}
\begin{array}{lll}
\E\left\{\phi(\ww R(\cdot),\ww a(\cdot),r(\cdot))\biggl|
        \Theta\right\}
& =&\E\left\{\Z_*\phi(\ww R_*(\cdot),\ww a_*(\cdot),r(\cdot))\biggl|
        \Theta\right\}\\
&=& \w\E\left\{\phi(\ww R_*(\cdot),\ww a_*(\cdot),r(\cdot))\biggl|
        \Theta\right\}\quad\hbox{a.s.}
\end{array}
\ee
Thus, for the next paragraph, without loss of generality, we will suppose that
$\Theta=\theta$ is deterministic, since for each value of $\Theta$ we can construct $\tilde R,\tilde R_*, \tilde a,\tilde a_*$ and $\w\P$.
\par
By Girsanov's Theorem, the process $$ \w w(t)\defi w(t)-\int_0^t
\a(s,\ww R_*(\cdot))^{-1}\ww a_*(s)ds $$ is a Wiener process under $\w\P$. From this and
(\ref{R}) we obtain
$$
\begin{array}{l}
d\ww R(t)=A(t,\Theta,\ww R(\cdot)|_{[0,t]})dt+\a(t,\ww R(\cdot)|_{[0,t]})dw(t),
\\
d\ww R_*(t)=A(t,\Theta,\ww R_*(\cdot)|_{[0,t]})dt+ \a(t,\ww R_*(\cdot)|_{[0,t]})d\w w(t).
\end{array}
$$
Then for each value of $\Theta$ the processes $\ww R(\cdot)$ and $\ww R_*(\cdot)$ have the same
distribution on the probability spaces defined by $\P$ and $\w\P$
respectively, and  (\ref{cond}), hence (\ref{phiphi2}) follows.

Further, (\ref{phiphi3}) follows by taking conditional expectation
in (\ref{phiphi2}). Finally, using Proposition~\ref{psi1} and
(\ref{phiphi2}),
\begin{eqnarray*}
\E_*\w\phi(\ww R(\cdot),r(\cdot))& =&\E \Z^{-1}
\w\phi(\ww R(\cdot),r(\cdot)) =\E \psi(\ww R(\cdot),\ww
a(\cdot))^{-1}
 \w\phi(\ww R(\cdot),r(\cdot))\\
 &=&\E \Z_*\psi(\ww R_*(\cdot),\ww a_*(\cdot))^{-1} \w\phi(\ww
R_*(\cdot),r(\cdot)) =\E \w\phi(\ww R_*(\cdot),r(\cdot)).
\end{eqnarray*} $\Box$ \par {\it Proof of Proposition
\ref{prop1}}. It suffices to show that $\E_*\phi\Z=\E_*\phi\oo \Z$
for all $\Fo_T$-measurable functions $\phi$. Such functions are of
the form $\w\phi(\ww R,r)$ above. But  (\ref{phiphi3}),
(\ref{phiphi4}) imply $$ \E_*\phi\Z=\E\phi=\E\w\phi(\ww R,r)=\E\oo
\Z_*\w\phi(\ww R_*,r)= \E_*\oo\Z\w\phi(\ww R,r)=\E_*\phi\oo\Z.$$
$\Box$ \par Remark 2.3 can be verified by a similar technique.
Just replace $\a(t,\cdot)$ by $\s(t,\o)$ no matter what the
argument in $\a$. From (\ref{vol}) it follows that there exists a
function ${\cal Q}$ such that $Q(t,\o)={\cal Q}(t,\ww
R(\cdot,\o))={\cal Q}(t,\ww R_*(\cdot,\o))$. If $A$ depends on $r$
also then $\oo\psi$ depends additionally on $r$. To obtain law
uniqueness we now condition on $\Theta,\s,r$ in the proof of
Proposition~\ref{propSS}.

We turn now to Theorem~\ref{ThM1}. Define $\w\xi_*\defi
F(\oo\Z_*,\w\lambda)$. If we define $\phi$ by $\w\xi=\phi(\ww
R(\cdot))$, then $\w\xi_*= \phi(\ww R_*(\cdot))$. \par {\it Proof
of Theorem \ref{ThM1}}. Let us show that $\E U^-(\w\xi)<\infty$ so
that $\E U(\w\xi)$ is well defined. For $k=1,2,...$, introduce the
random events $$ \O_*^{(k)}\defi\bigl\{-k\le U(\w\xi_*)\le
0\bigr\}, \quad \O^{(k)}\defi\bigl\{-k\le U(\w\xi)\le 0\bigr\}, $$
along with their indicator functions, $\chi_*^{(k)}$ and
$\chi^{(k)}$, respectively. The number $\w\xi_*$  provides the
unique  maximum of the function $\oo\Z_* U(\xi_*)-\w\lambda \xi_*$
over $\w D$, and $ X_0\in \w D$. Hence by
Proposition~\ref{propSS}, we have, for all $k=1,2,...$, $$
\begin{array}{rl} \E \chi^{(k)}
U(\w\xi)-\E\chi_*^{(k)}\w\lambda\w\xi_* &=\E
\chi_*^{(k)}\left(\oo\Z_* U(\w\xi_*)-\w\lambda\w\xi_*\right) \ge
\E\chi_*^{(k)}\left(\oo\Z_* U(X_0)-\w\lambda X_0\right) \\ &= \E
\chi^{(k)}U(X_0)-\w\lambda X_0\P(\O_*^{(k)}) \ge
-|U(X_0)|-|\w\lambda X_0|>-\infty. \end{array} $$ Furthermore, we
have that $\E|\w\xi_*|=\E_*|\w\xi|<+\infty$. Hence $\E
U^-(\w\xi)<\infty$. \par Now observe that for any $\pi\in{\cal A}$
we can apply (\ref{phiphi3}) and (\ref{phiphi4}) to $U(\ww
X^\pi(T))$ (and use (\ref{lZb})) to obtain $$\ba \E U(\ww
X^\pi(T))&=\E_*\{\oo\Z U(\ww X^\pi(T))\}\le\E_*\{\oo\Z U(\ww
X^\pi(T))-\w\lambda \ww X^\pi(T)\}
        +\w\lambda X_0\\ &\le\E_*\{\oo\Z U(\w\xi)-\w\lambda
\w\xi\} +\w\lambda X_0=\E_* \oo\Z U(\w\xi)= \E U(\w\xi). \ea $$
Thus (ii) is satisfied. \par To show (iii), note that ${\cal
F}^w_t={\cal F}^{\ww R_*}_t$ so $\w\xi_*=\phi(w(\cdot))$, where
$\phi(\cdot):B([0,T];{\bf R}^n)\to {\bf R}$ is a measurable
function. By the martingale representation theorem, $$
\w\xi_*=\E\w\xi_*+\int_0^T f(t,w(\cdot)|_{[0,t]})^\top dw(t), $$
where $f(t,\cdot):B([0,t];{\bf R}^n)\to {\bf R}$ is a measurable
function such that $
\int_0^T|f(t,w(\cdot)|_{[0,t]})|^2dt<+\infty\quad \hbox{a.s.} $
There exists  a unique  measurable function
$f_0(t,\cdot):B([0,t];{\bf R}^n)\to {\bf R}$ such that
$f(t,w(\cdot)|_{[0,t]})\equiv f_0(t,\ww R_*(\cdot)|_{[0,t]})$.
Thus, \be  \label{xh} \w\xi_*=\E\w\xi_*+\int_0^Tf_0(t,\ww
R_*(\cdot)|_{[0,t]})^\top dw(t)= \E\w\xi_*+\int_0^Tf_0(t,\ww
R_*(\cdot)|_{[0,t]})^\top \a(t,\ww R_*)^{-1}d\ww R_*(t). \ee
Proposition~\ref{propSS} implies that $\E\w\xi_*= \E_*\w\xi=X_0$,
and $$ \w\xi=X_0 +\int_0^T f_0(t,\ww R(\cdot)|_{[0,t]})^\top
\s(t)^{-1}d\ww R(t). $$ It follows that  the  strategy
$\w\pi(t)^\top=B(t)f_0(t,\ww R(\cdot)|_{[0,t]})^\top \s(t)^{-1}$
replicates $B(T)\w\xi$. It  belongs to ${\cal A}$; in
particular,  $\ww X^{\w\pi}(t)=X_0+\int_0^tf_0(t,\ww
R(\cdot)|_{[0,t]})^\top \s(t)^{-1}d\ww R(t)=\E_*(\w\xi\,|\F^{\ww
R}_t)\in\w D$ since $\w D$ is convex. Hence $\ww X^{\w\pi}$ is
bounded below. This completes the proof of Theorem \ref{ThM1}.
$\Box$ \par  {\it  Proof of Corollary \ref{corlog}(ii)}.
  We shall employ the notation
$Y(t,\pi)\defi\log\left(\frac{\ww X^\pi(t)+\delta}{X_0+\delta}\right)$.
Let ${\cal B}_2$ be the set of all processes $\oo a(t):[0,T]\to{\bf R}^n$
which are progressively measurable with respect to $\Fo_t$
and such that $\E\int_0^T|\oo a(t)|^2dt<+\infty$.
For any $\oo a(\cdot)\in{\cal B}_2 $, define $\oo\pi(t)^\top\defi
(X^{\oo\pi}(t)+\delta B(t))\oo a (t)^\top Q(t)$ where
 $X^{\oo\pi}(t)\defi B(t)\ww X(t)$ and $\ww X(\cdot)$ is found from
 (\ref{nw}) using $\pi^\top=B (\ww X +\delta) \oo a^\top Q$. Then
\be \label{YY}
Y(t,\oo\pi)=
\int_0^t  \left(\oo a(s)^\top Q(s)
\,d\ww R(s)
-\frac{1}{2}\int_0^t \oo a(s)^\top Q(s)\oo a(s)ds\right),
\ee
and
\be\label{Y'}\E Y(T,\oo\pi)=\frac{1}{2}\E
\int_0^T\left(- |\s(t)^{-1}(\oo a(t)-\ww a(t))|^2
+\ww a(t)^\top Q(t)\ww a(t)\right)dt.
\ee\par
Set  $a'(t)\defi\E\{\ww a(t)|\Fo_t\}$.
Since $\E |K(\Theta,\s(\cdot),\s(\cdot)^\top,r(\cdot))|^2<\infty$, then
 Jensen's inequality implies that  $a'(\cdot)\in{\cal B}_2$.
Consider the corresponding strategy
\be
\label{s'}
\pi'(t)^\top\defi  (X^{\pi'}(t)+\delta B(t))a' (t)^\top Q(t).
\ee
 It is well known that
$\E Y(T,\pi')\ge \E Y(T,\oo\pi)$,  so the strategy (\ref{s'}) is
optimal over all $\oo\pi(\cdot)$ which correspond to $\oo
a(\cdot)\in{\cal B}_2$. Then (\ref{EU}) and the Corollary follow
if $\w a(\cdot)\in{\cal B}_2$.
\par
Let us show this. For any $K>0$, set
$$
T_K\defi\inf\{t\in [0,T]:\quad
\int_0^t|\w a(s)|^2ds>\int_0^t|a'(s)|^2ds+K\}.
$$
As usual we take $T_K=T$ if the set is empty.
Note that
\be
\label{Mark}
\E \log (\ww X^{\w \pi}(T_K)+\delta)\ge\E \log (\ww X^{\pi'}(T_K)+\delta)
\quad \forall K>0,
\ee
because if (\ref{Mark}) fails, then
$\E Y(T,\pi_K)>\E Y(T,\w \pi)$,
where
$$
\pi_K(t)\defi\cases{\pi'(t) & $t\le T_K$\cr
\w \pi(t) & $t> T_K.$\cr}
$$
Further, let $\chi_K(t)$ denote the indicator function
of the event $\{t<T_K\}$ and let $\oo a_K(\cdot)\defi\chi_K(\cdot)\oo a(\cdot)\in{\cal B}_2,\ \ww a_K(t)\defi\chi_K(t)\ww a(t)$. As in (\ref{Y'}), we have
\begin{eqnarray}
\label{YMark}
\E Y(T_K,\oo \pi)
&=& \frac{1}{2}
\E\int_0^{T_K}\left(-|\s(t)^{-1}( \oo a(t)-\ww a(t))|^2
+\ww a(t)^\top Q(t)\ww a(t)\right)dt\\
&=&\frac{1}{2}
\E\int_0^{T}\left(-|\s(t)^{-1}(\oo a_K(t)
-\ww a_K(t))|^2
+\ww a_K(t)^\top Q(t)\ww a_K(t)\right)dt.\nonumber
\end{eqnarray}
Then the process
$\E\{\ww a_K(t)|\Fo_{t}\}=\chi_K(t)\E\{\ww a(t)|\Fo_{t}\}=\chi_K(t)a'(t)$
 gives the maximum
of $\E Y(T_K,\oo \pi)$.
It follows from (\ref{Mark}) that
$\chi_K(t)\w a(t)=\chi_K(t)a'(t)$ for $t\in[0,T]$ and $K>0$.
Thus, $T_K=T$ a.s. for any $K>0$,
and $a'(\cdot)=\w a(\cdot)$, $\w a(\cdot)\in {\cal B}_2$.
Then (\ref{EU}) and (ii) follow.
 $\Box$ \par  {\it Proof of Theorem \ref{Thm}}. From
Conditions~\ref{ass01} and \ref{assla} it follows that
$F(z,\lambda)=\lambda^{-l} z^l$, so we want to replicate
$B(T)(X_0/\E_*\oo\Z^l)\oo\Z^l$. Let us first find a
representation for $\oo\Z^l$. $$ \begin{array}{lll} \oo\Z^l&=&
\int_{{\cal T}^{l}} d\nu(\theta_1)\cdots
d\nu(\theta_{l})\exp\biggl(\sum_{k=1}^{l}\int_0^T
\theta_k(t)^\top Q(t) d\ww R(t)
-\frac{1}{2}\sum_{k=1}^{l}\int_0^T \theta_k(t)^\top
Q(t)\theta_k(t)dt\biggr)\\ & =& \int_{{\cal T}^{l}}
d\nu(\theta_1)\cdots d\nu(\theta_{l})\g(\theta_1,...,\theta_l)
z\left(\sum_{k=1}^{l}\theta_k,T\right)\\ &=& \int_{\oo\T}
d\oo\nu(\theta)z(\theta,T) G. \end{array} $$ It follows that
$\E_*\,\oo\Z^l=G$ since $\E_*z(\theta,T)=\E_*\psi(\ww
R,\theta)=\E\psi(\ww R_*,\theta)=1$ because $\psi(\ww
R_*,\theta)$ is a $\P$ martingale. \par Now we must show that
$\ww X^{\w\pi}(T)=X_0\oo\Z^l/G$. But $$
\frac{X_0\oo\Z^l}{G}=X_0\int_{\oo\T}d\oo\nu(\theta)z(\theta,T)=X_0\left(1+\int_0^T\int_{\oo\T}d\oo\nu(\theta)z(\theta,t)\theta(t)^\top
Q(t)\,d\ww R(t)\right)=\ww X^{\w\pi}(T) $$ with $\w\pi$ defined
by (\ref{powpi}). \par

Now (\ref{powX}) and the equality in (\ref{powpi}) follow from
(\ref{nw}) and (\ref{za}). Finally (\ref{Ezm=2}) follows from
$\E\{\oo \Z^{l-1}\}=\E_*\{\oo \Z^l\}=G$. $\Box$
\par
{\it Proof of Corollary \ref{El}}.
If we take $\T=\{\theta_o\}$ then $\oo\T=\{l\theta_o\}$, so (\ref{powpi}) implies that the optimal strategy in case of complete observation is  $\pi(t)=lX^\pi(t)Q(t)\ww a(t)$, hence the
first equality in (\ref{estm}) follows. This and (\ref{powpi}), (\ref{powX}) imply
$$
l \w a(t)^\top=\frac{\int_{\oo\T} d\oo\nu(\theta) z( \theta,t) \theta(t)^\top}{\int_{\oo\T} d\oo\nu(\theta) z( \theta,t)}.
$$
Comparing this with Corollary \ref{corlog}(i) and (\ref{logX}),
we see that $l\w a(t)$ is the
equivalence filter for the problem with
$U(x)\equiv\log x$ and with the prior distribution of $\Theta = \ww a(\cdot)$
described by $\oo\nu$ on $\oo\T$.
By Corollary \ref{corlog} (ii),  $\w a(t)=l^{-1}\oo\E \{\ww a(t)|\Fo_t\}$.
$\Box$
\par
{\it Proof of Proposition \ref{prop01}}. It is required to show that the strategy defined in
the Proposition does exists and  is admissible.
Assume that  $C(\cdot)$ has a finite support inside an open domain
in ${\bf R}^M$, and let the function $C(\cdot)$ be smooth enough. Set $V(x,s)\defi \E_*C(y^{x,s}(T))$, where $y^{x,s}(\cdot)$ is the solution of \be
\label{yMark2}
 \left\{
\begin{array}{ll}
dy(t)=f(y(t),t)dt+\oo b(y(t),t)d\ww R(t),\\
y(s)=x.\end{array}\right. \ee
Then it can be shown that $V(x,s)$ is the classical solution of
the problem (\ref{parab1})-(\ref{parab1'}). Thus,
$V(x,t)$ is a classical solution of (\ref{parab1})-(\ref{parab1'}).
Set $\ww X_*(t)=V(y_*(t),t)$. From (\ref{parab1})  and
 It\^o's Lemma, it follows that
$$
\ww X_*(T)=\ww X_*(t)+\int_t^T \frac{\p V}{\p y}(y_*(s),s)\oo b(y_*(s),s)\,dw(s).
$$
It follows that $\ww X_*(0)=V(y_*(0),0)=\E V(y_*(T),T)=X_0$
and
\be
\label{2} d\ww X_*(t)=\frac{\p V}{\p y}(y_*(t),t)^\top\oo b(y_*(t),t) d\ww R_*(t),\quad
\ww X_*(T)=C(y_*(T)).
\ee
Then $\ww X_*(t)=\psi(t,\ww R_*)$ for some measurable $\psi$, and the result
follows if we observe that
 $\ww X^\pi(t)=\psi(t,\ww R)$ for the given $\pi$.
\par
To continue, we require some a priori estimates. Let $\zeta(t)\defi
\a(t,\ww R(t))^\top\pi(t)$. Define $\pi_*$ in the obvious way.  Consider the conditional probability space
given $r(\cdot)$. With respect to the conditional probability
space, it follows from (\ref{2})  that
\be
\label{4} \left\{ \begin{array}{ll} d\ww X_*(t)=B(t)^{-1}\zeta_*(t)^\top
dw(t),
\\
 \ww X_*(T)=C(y_*(T)).
\end{array}
\right.
 \ee The solution $(Z_*(t),\ww X_*(t))$ of the stochastic backward
equation (\ref{4})
 is a process in $L_2([0,T],L^2(\O,\F,P))\times C([0,T],L^2(\O,\F,P))$
(see, e.g.. El Karoui {\it et al} (1997), or Yong and Zhou (1999),
Chapter 7, Theorem 2.2). Note that the equation (\ref{4}) is
linear. Thus, it can be shown by using Theorem 2.2 from Chapter 7
from Yong and Zhou (1999) again
  that
there exists a constant $c_0$, independent of $C(\cdot)$, such
that
 $$
\begin{array}{ll} \sup_t\E\left\{ |\ww X_*(t)|^2\bigl|\,
r(\cdot)\, \right\}+ \E\biggl\{\int_0^T|\zeta_*(t)|^2dt&\biggl|\,
r(\cdot)\,\biggr\} \le c_0\E\left\{C(y_*(T))^2\bigl|\, r(\cdot)\,
\right\}\quad\hbox{a.s}.
\end{array}
$$
 Hence
\be
\label{est1} \sup_t\E |\ww X_*(t)|^2+ \E\int_0^T|\zeta_*(t)|^2dt \le
c_0\E C(y_*(T))^2. \ee
\par
Let $C(\cdot)$ be a general measurable function satisfying the
conditions specified in the proposition. Then, there exists a
sequence
 $\{C^{(i)}(\cdot)\}$, where $C^{(i)}(\cdot)$ has a
 finite support   inside the open domain
${\bf R}^M$ and is smooth enough,
 such that
$$ \E |C^{(i)}(y_*(T))-C(y_*(T))|^2\to 0\quad\hbox{as}\quad
i\to\infty. $$
 Let $\ww X_*^{(i)}(\cdot)$, $\pi_*^{(i)}(\cdot)$,
 $V^{(i)}(\cdot)$
be the corresponding processes and functions.
By (\ref{est1}) and the linearity of (\ref{4}), it follows that $$
\begin{array}{ll}
\sup_t\E |\ww X_*^{(i)}(t)&-\ww X_*^{(j)}(t)|^2+ \E\int_0^T
|\pi_*^{(i)}(t)-\pi_*^{(j)}(t))|^2dt
\\
&\le c_0\E |C^{(i)}(y_*(T))-C^{(j)}(y_*(T))|^2\to 0
\quad\hbox{as}\quad i\to\infty.
\end{array}
$$
From (\ref{phiphi4}) it follows that
$$
\sup_t\E_* |\ww X^{(i)}(t)-\ww X^{(j)}(t)|^2+ \E_*\int_0^T
|\pi^{(i)}(t)-\pi^{(j)}(t))|^2dt\to 0
\quad\hbox{as}\quad i\to\infty.
$$
Thus,   $\{\ww X^{(i)}(\cdot)\}$, $\{\pi^{(i)}(\cdot)\}$ are Cauchy sequences in the
space the spaces $C([0,T],L^2(\O,\F,P_*))$ and $L_2([0,T],L^2(\O,\F,P_*))$
correspondingly, and hence, it can be shown that the corresponding limits $\ww
X(\cdot)$, $\pi(\cdot)$  exist, and belongs these spaces.  Similarly Dokuchaev and
Zhou (2001), it follows from the definition of $\Y^1$ that $V^{(i)}(\cdot)$ is a
 Cauchy sequence in and $\Y^1$. This completes the proof. $\Box$
\par {\it Proof of Theorem \ref{ThMarkov}.} As in the proof
above, it can be shown that  $\ww X(t)=V(y(t),t,\w\lambda)$ is
the solution of some equation (\ref{2}), i.e. it is the
normalized wealth. Then the proof follows. $\Box$ \vspace{0mm}
\subsection*{Acknowledgment}  This work  was supported by  Australian Research Council grant  DP120100928 to the author.
\section*{References} $\hphantom{xk}$ Brennan, M.J. (1998): { The
role of learning in dynamic portfolio decisions.} {\em European
Finance Review} {\bf 1}, 295--306. \par Cvitani\'c, J., Ali
Lazrak,  Martellini, L., and Zapatero, F. (2002): A note on
portfolio selection with partial information: power utility and
Gaussian prior. Working paper. \par Chen, R.-R., and Scott, L.
(1993): { Maximum likelihood estimation for a multifactor
equilibrium model of the term structure of interest rates}. {\em
Journal  of Fixed Income} {\bf 4}, 14--31.
 \par
Detempte, J.B. (1986): { Asset pricing in an economy with incomplete information.}
{\em Journal  of Finance} {\bf 41}, 369--382.
\par
Dokuchaev, N.G., and Haussmann, U. (2001): Optimal portfolio selection and compression
in an incomplete market.  {\em Quantitative Finance} {\bf 1} (3), 336--345.
\par  Dokuchaev, N.G., and Teo, K.L. (2000): Optimal hedging strategy for a portfolio
investment problem with  additional constraints. {\em Dynamics of Continuous, Discrete
and Impulsive Systems} {\bf 7}, 385--404.
\par
{ Dokuchaev, N.G., and Zhou, X.Y.} (2001): Optimal investment strategies with bounded
risks, general utilities, and goal achieving.  {\em Journal  of Mathematical
Economics} {\bf 35} (2), 289--309.
\par
 Dokuchaev, N.G., and Savkin, A.V. (2002). A bounded risk
strategy for a market with non-observable parameters.  {\em
Insurance: Mathematics and Economics}, {\bf 30}, 243-254.
  Dokuchaev N.G.(2002) {\it Dynamic portfolio
strategies: quantitative methods and empirical rules for incomplete
information.} Kluwer Academic Publishers, Boston.
 \par
Dokuchaev, N.G. (2005). Optimal solution of investment
problems via linear parabolic equations generated by Kalman filter.
{\it SIAM J. of Control and Optimization} {\bf 44},  No. 4, pp.
1239-1258. 
\par
 Dokuchaev, N. (2007).
Mean-reverting market model: speculative opportunities and
  non-arbitrage.  {\it   Applied Mathematical Finance} {\bf 14},
  iss. 4, 319-337.
  \par
Dothan, U., and Feldman, D. (1986): { Equilibrium interest rates and multiperiod bonds
in a partially observable economy.} {\em Journal  of Finance} {\bf 41}, 369--382.
\par El
Karoui, N., Peng, S.,  and Quenez, M.C. (1997): {Backward stochastic differential
equations in finance.} {\em Mathematical Finance} {\bf 7},  1--71.
\par
Feldman, D. (2007). Incomplete information equilibria: Separation theorems and other myths.
{\em Annals of Operations Research},
{\bf 151}, Iss. 1, pp. 119--149. 
    \par
Frittelli, M. (2000): {Introduction to a theory of value coherent with the
no-arbitrage  principle}, {\it Finance and Stochastics}, {\bf 3}, 275-298.
\par
Gennotte, G. (1986): { Optimal portfolio choice under incomplete information}. {\em
Journal of Finance} {\bf 41}, 733--749.
\par  Hakansson, N.H. (1971): { Multi-period mean-variance
analysis: Toward a general theory of portfolio choice.} {\em Journal of  Finance} {\bf
26}, 857--884.
\par Karatzas, I. (1997): { Adaptive control of a
diffusion to a goal and a parabolic  Monge--Amp\'ere type equation}.  {\em The Asian
{Journal } of Mathematics} {\bf 1}, 324--341.
\par
Karatzas, I., and   Shreve, S.E. (1991): {\em Brownian Motion and Stochastic
Calculus}, 2nd edn. Berlin, Heidelberg, New York: Springer.
\par
 Karatzas, I., and   Shreve, S.E. (1998):  {\em Methods of Mathematical
Finance}. New York: Springer-Verlag.
\par   Karatzas, I., and Zhao, X. (2001): { Bayesian adaptive portfolio optimization}.
 In: Cvitani\'c, J. et al. (eds.), {\it Handbook of  Mathematical Finance},
 Cambridge University Press, pp. 632-669.
\par
Kuwana, Y. (1995): { Certainty equivalence and logarithmic utilities in consumption/
investment problem.} {\em Mathematical Finance} {\bf 5}, 297--310.
\par
Lakner, P. (1995): Utility maximization with partial information. {\em Stoch.
Processes Appl.} {\bf 56}, 247-273.
\par
Lakner,P,  (1998): Optimal trading strategy for an investor: the case of partial
information. {\em Stochastic Processes and their Applications}, {\bf 76}, 77-97.
\par
\par
Liptser,R.S., and  Shiryaev, A.N. {\em Statistics of Random Processes. I. General
Theory}, Springer-Verlag. Berlin, Heidelberg, New York, 2nd edn.   2000.
\par
Lo, A.W. (1998): {Maximum likelihood estimation of generalized It\^o processes with
discretely sampled data.} {\em Econometrics Theory} {\bf 4}, 231--247.
\par
Merton, R. (1969):  Lifetime portfolio selection under uncertainty: the
continuous-time case.  {\em Review of Economics and Statistics} {\bf 51}, 247--257.
\par
 Pearson, N.D.,
and Sun, T.-S. (1994): { Exploiting the conditional density in estimating the term
structure: An application to the Cox, Ingresoll,
 and Ross model}. {\em Journal  of Finance} {\bf 49}, 1279--1304.
 \par
Sass, J., and Haussmann, U. (2003): Optimizing the terminal wealth
under partial information: the drift process as a continuous time
Markov chain. Preprint.
\par Williams,
J.T. (1977): {Capital assets prices with heterogeneous beliefs.}
{\em Journal  of Financial Economics} {\bf 5}, 219--240.
\par
Yong, J., and Zhou, X. Y. (1999): {\em  Stochastic Controls: Hamiltonian Systems and
HJB Equations.} Springer-Verlag.  New York.
\par
Zohar, G. (2001): Generalized Cameron-Martin Formula with
applications to Partially Observed Dynamic Portfolio Optimization,
{\em Mathematical Finance} {\bf 11},  475-494.
\end{document}